\documentclass[fleqn,10pt]{SelfArx} 

\definecolor{color1}{RGB}{0,0,90} 
\definecolor{color2}{RGB}{0,20,20} 

\usepackage{hyperref} 
\hypersetup{hidelinks,colorlinks,breaklinks=true,urlcolor=color2,citecolor=color1,linkcolor=color1,bookmarksopen=false,pdftitle={Title},pdfauthor={Author},urlcolor=blue}

\usepackage{graphicx}
\usepackage[square]{natbib}
\usepackage{amsmath}
\usepackage{multirow}
\usepackage{subfigure}
\usepackage{pdflscape}

\newcommand{\n}[1]{\mathrm{#1}}

\JournalInfo{Published in Journal of Magnetism and Magnetic Materials, Vol. 384, 128–132, 2015} 
\Archive{\href{http://dx.doi.org/10.1016/j.jmmm.2015.02.034}{DOI: 10.1016/j.jmmm.2015.02.034}} 

\PaperTitle{The efficiency and the demagnetization field of a general Halbach cylinder} 

\Authors{R. Bj\o{}rk, A. Smith and C. R. H. Bahl} 
\affiliation{\textit{Department of Energy Conversion and Storage, Technical University of Denmark - DTU, Frederiksborgvej 399, DK-4000 Roskilde, Denmark}} 
\affiliation{*\textbf{Corresponding author}: rabj@dtu.dk} 

\newcommand{\keywordname}{Keywords} 

\Abstract{The maximum magnetic efficiency of a general multipole Halbach cylinder of order $p$ is found as function of $p$. The efficiency is shown to decrease for increasing absolute value of $p$. The optimal ratio between the inner and outer radius, i.e. the ratio resulting in the most efficient design, is also found as function of $p$ and is shown to tend towards smaller and smaller magnet sizes. Finally, the demagnetizing field in a general $p$-Halbach cylinder is calculated, and it is shown that demagnetization is largest either at $\cos 2p\phi=1$ or $\cos 2p\phi=-1$. For the common case of a $p=1$ Halbach cylinder the maximum values of the demagnetizing field is either at $\phi = 0,\pi$ at the outer radius, where the field is always equal to the remanence, or at $\phi = \pm \pi/2$ at the inner radius, where it is the magnitude of the field in the bore. Thus to avoid demagnetization the coercivity of the magnets must be larger than these values.}

\begin{document}

\flushbottom 

\maketitle 


\thispagestyle{empty} 

\section{Introduction}
Generating a powerful magnetic field with a permanent magnet assembly is important in a number of applications. It is also often of importance that the magnetic field is generated using the least amount of magnet material possible. A common magnet design used to generate such powerful magnetic fields is the Halbach cylinder, which has been used in a number of applications, such as nuclear magnetic resonance (NMR) equipment \cite{Moresi_2003,Appelt_2006}, accelerator magnets \cite{Sullivan_1998,Lim_2005}, magnetic refrigeration devices \cite{Tura_2007,Bjoerk_2010b} and medical applications \cite{Sarwar_2012}. The Halbach cylinder is a hollow cylinder made of a ferromagnetic material with a remanent flux density which in cylindrical coordinates is given by
\begin{align}
B_{\mathrm{rem},r}    &= B_{\mathrm{rem}}\; \cos p\phi \nonumber\\
B_{\mathrm{rem},\phi} &= B_{\mathrm{rem}}\; \sin p\phi,\label{Eq.Halbach_magnetization}
\end{align}
where $B_{\mathrm{rem}}$ is the magnitude of the remanent flux density and $p$ is an integer \cite{Mallinson_1973,Halbach_1980}. For $p$ positive a field inside the cylinder bore is generated, which for the important case of $p=1$ is spatially uniform. There is zero field outside the cylinder. For $p$ negative the Halbach cylinder creates a field on its outside, while the inside field becomes zero. Usually, as will also be the case here, a two dimensional problem is considered. This is a good approximation if the radius of the cylinder is smaller that its length. The magnetic field distribution for such a Halbach cylinder of infinite length \cite{Zhu_1993,Atallah_1997,Peng_2003,Xia_2004,Bjoerk_2010a} as well as for finite length \cite{Mhiochain_1999,Xu_2004,Bjoerk_2008,Bjoerk_2011b} have previously been investigated in detail. However, the efficiency of Halbach cylinders have not been considered for a general $p$-Halbach cylinder.

The plan of the paper is as follows: First we calculate the efficiency of a $p$-Halbach cylinder and thereafter we consider the demagnetization field internally in the Halbach cylinder and its possible influence on the performance and efficiency of the Halbach cylinder. Finally, we discuss the implications of our findings.

\begin{figure}[t]
  \centering
  \includegraphics[width=0.8\columnwidth]{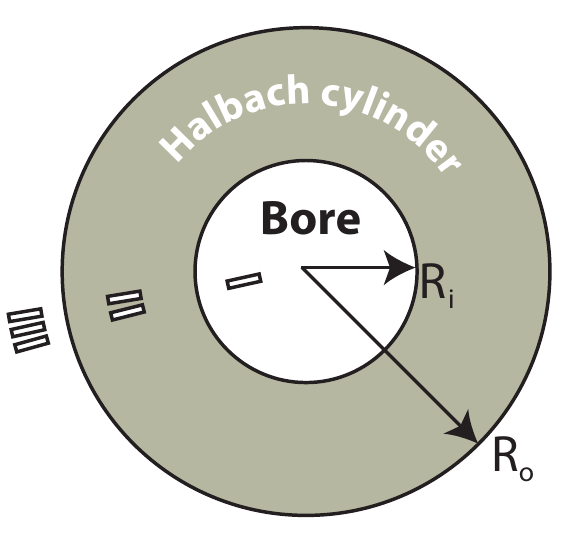}
  \caption{The Halbach cylinder geometry for a $p$-Halbach ($p>0$). The different radii and regions, I-III, have been indicated.}
  \label{Fig_Combined_drawing}
\end{figure}

We consider the field from a $p$-Halbach in general. The geometry is as shown in Fig.~\ref{Fig_Combined_drawing} for the case of an `interior' Halbach (positive $p$); for an `exterior' Halbachs (corresponding to negative $p$) the field is generated on the outside of the cylinder. We assume that the permanent magnets are perfectly linear and with a relative permeability of $\mu_\textrm{r}=$1, which is a good approximation for NdFeB magnets.

\section{Efficiency of a Halbach cylinder}
The object of a permanent magnet array is to generate a magnetic field of given characteristics in a given volume. Many different magnet configurations can in principle produce the same magnetic field, and thus the question arises of how to do it most efficiently. Jensen and Abele \cite{Jensen_1996} proposed a general figure of merit, $M^{*}$, to characterize the efficiency of a given magnet design:
\begin{equation}\label{Eq.Mstar_definition}
M^{*}=\frac{\int_{V_\n{field}}||\mathbf{B}||^2dV}{\int_{V_\n{mag}}||\mathbf{B_\n{rem}}||^2dV},
\end{equation}
where $V_\n{field}$ is the volume of the region where the magnetic field is created and $V_\n{mag}$ is the volume of the magnets. The figure of merit is the ratio of the energy stored in the field region to the maximum amount of magnetic energy available in the magnetic material. It can be shown that the maximum value of $M^{*}$ is 0.25 \cite{Jensen_1996}.

We here consider the efficiency of a $p$-Halbach cylinder. While the efficiency of a $p=1$ Halbach cylinder have been considered several times \cite{Bjoerk_2011b,Abele_1990,Coey_2003}, the efficiency of a general $p\neq{}1$ Halbach cylinder have not been considered in the literature, although the calculation is straight-forward. The efficiency can be found using the analytical solution to the field equations for the magnetic field for a general $p$-Halbach cylinder. These are given in Bj\o{}rk et. al. \cite{Bjoerk_2010a}, where the field interior to the cylinder is given as:
\begin{equation}\label{interiorfield}
  \begin{pmatrix} B_r(r,\phi) \\ B_\phi(r,\phi) \end{pmatrix} = \begin{cases} B_\n{rem}\frac{p}{p-1}\left[1- \left(\frac{R_i}{R_o}\right)^{p-1}\right]\left(\frac{r}{R_i}\right)^{p-1} \left(\begin{smallmatrix} \cos p\phi \\ -\sin p\phi \end{smallmatrix}\right) & p > 1 \\
  B_\n{rem}\ln\left(\frac{R_o}{R_i}\right)
  \left(\begin{smallmatrix} \cos \phi \\ -\sin \phi \end{smallmatrix}\right) & p=1. \end{cases}
\end{equation}
Here $R_o$ is the outer radius of the Halbach and $R_i$ the inner radius, as shown in  Fig. \ref{Fig_Combined_drawing}.

For an exterior Halbach ($p\leq -1$) the field outside the Halbach is:
\begin{eqnarray}\label{exteriorfield}
  \begin{split}\begin{pmatrix} B_r(r,\phi) \\ B_\phi(r,\phi) \end{pmatrix} = & B_\n{rem}\frac{p}{p-1} \left[1- \left(\frac{R_o}{R_i}\right)^{p-1}\right]*\\
  &\left(\frac{r}{R_o}\right)^{p-1} \begin{pmatrix} \cos p\phi \\ -\sin p\phi \end{pmatrix}\end{split}
\end{eqnarray}

Performing the integral in Eq.~(\ref{Eq.Mstar_definition}) for an internal and external Halbach cylinder yields
\begin{equation}\label{Eq.MstarHalbach}
M^{*} = \begin{cases} \frac{(R_i/R_o)^2}{1-(R_i/R_o)^2}\frac{p}{(1-p)^2}\left(1-\left(\frac{R_i}{R_o}\right)^{p-1}\right)^2 & p > 1 \\
  \ln\left(\frac{R_o}{R_i}\right)^2 \frac{(R_i/R_o)^2}{1-(R_i/R_o)^2} & p=1
  \\ -\left(\frac{R_i}{R_o}\right)^{-2p} \frac{(R_i/R_o)^2}{1-(R_i/R_o)^2}\frac{p}{(1-p)^2}\left(1-\left(\frac{R_i}{R_o}\right)^{p-1}\right)^2 & p\leq -1 \end{cases}
\end{equation}

\begin{figure}[t]
  \centering
  \includegraphics[width=0.8\columnwidth]{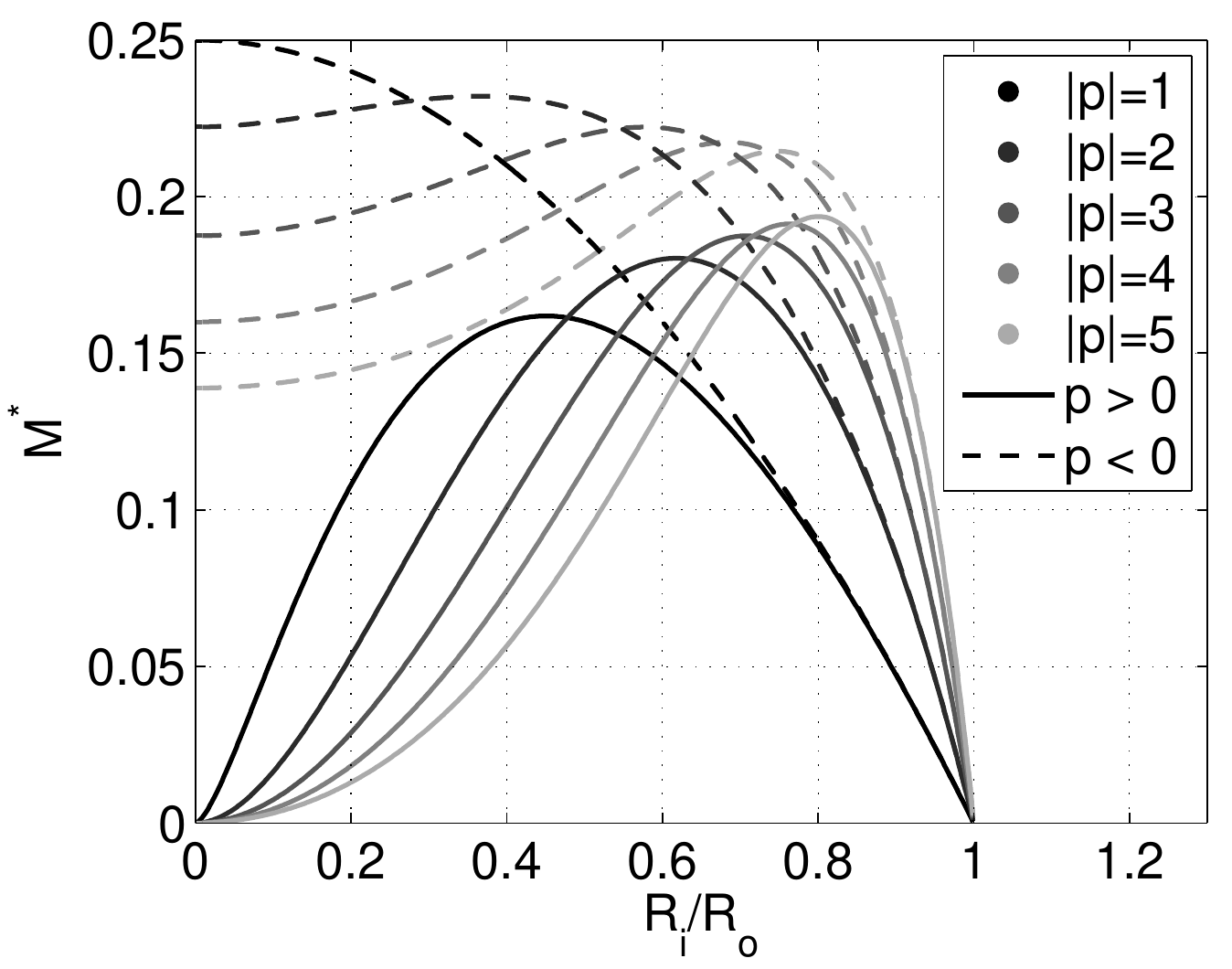}
  \caption{The efficiency, $M^{*}$, as function of the ratio of the inner and outer radius for Halbach cylinders with $p=-5$ to $p=5$.}
  \label{Fig_Mstar_p}
\end{figure}

The efficiency as a function of the ratio between the radii is plotted in Fig. \ref{Fig_Mstar_p} for $p=-5$ to $p=5$.

\begin{figure}[t]
  \centering
  \includegraphics[width=0.8\columnwidth]{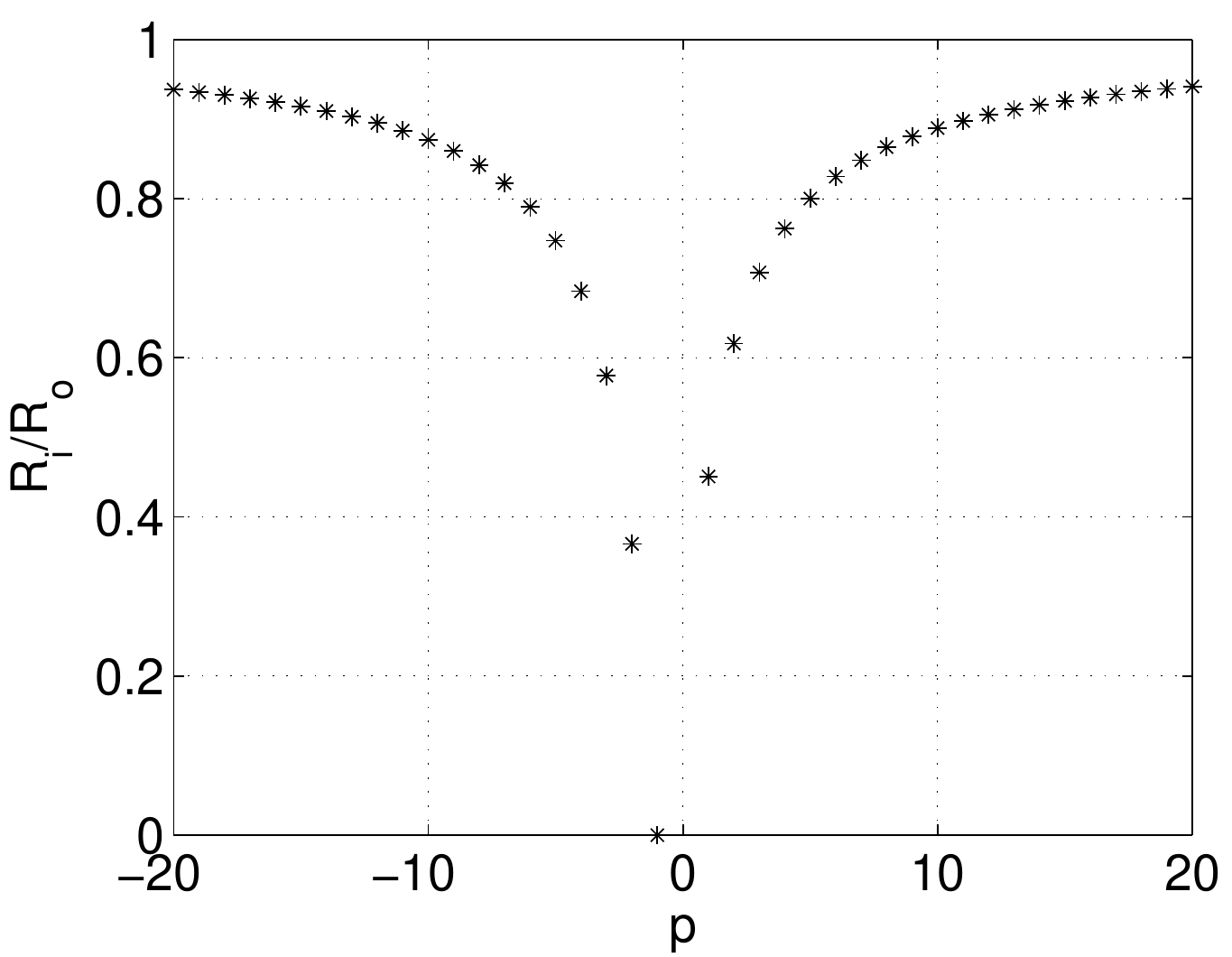}
  \caption{The optimal ratio between the inner and outer radius, i.e. the ratio producing the highest efficiency $M^{*}$, as function of $p$.}
  \label{Fig_Optimal_ratio_p}
\end{figure}

\begin{figure}[t]
  \centering
  \includegraphics[width=0.8\columnwidth]{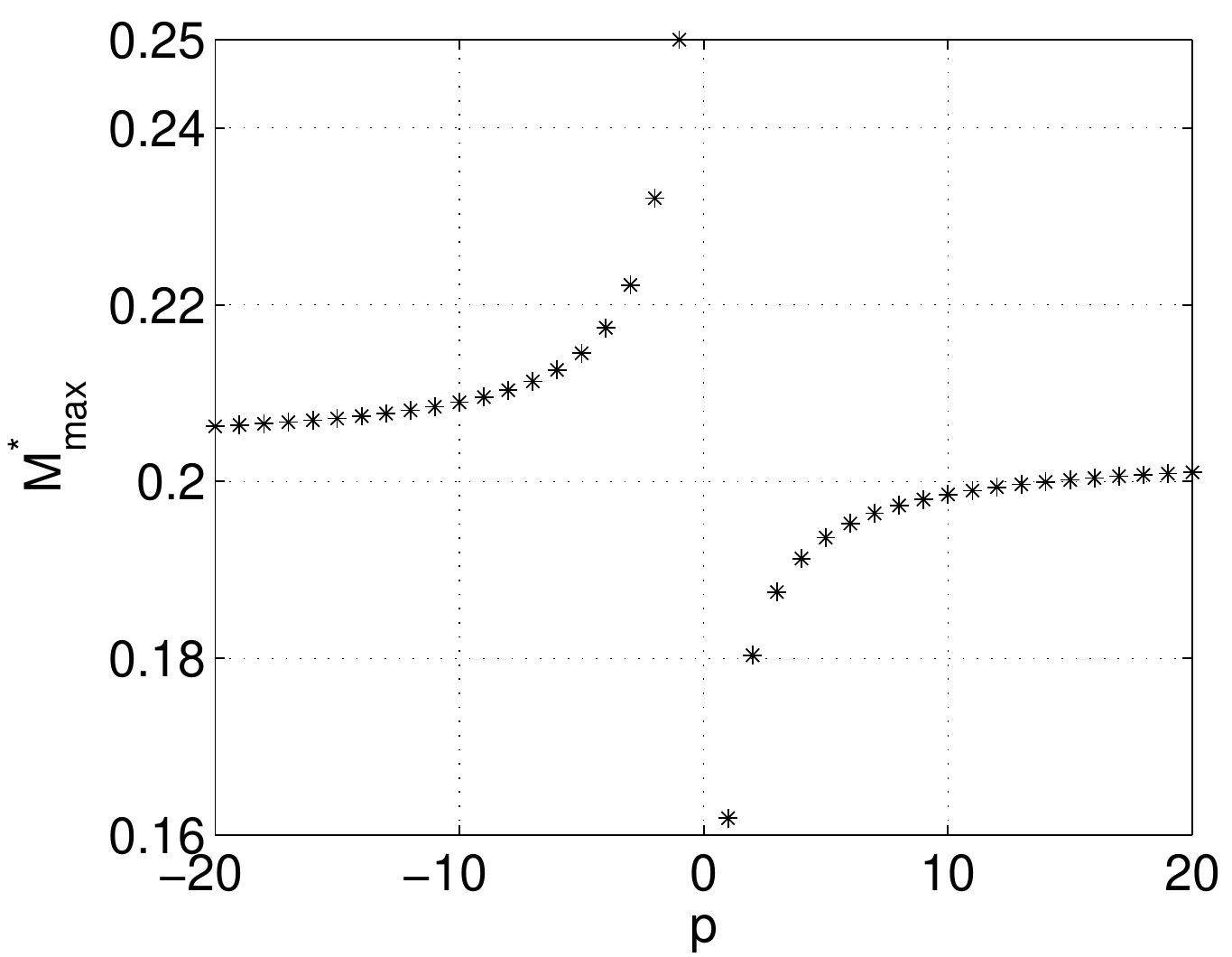}
  \caption{The highest value of the efficiency for a $p$ Halbach cylinder as function of $p$.}
  \label{Fig_Maximum_Mstar_p}
\end{figure}

It is of interest to determine the optimal efficiency possible for a general Halbach cylinder and the ratio of the inner and outer radius at which this occurs. This can be done by taking the derivative of $M^*$ (Eq.~(\ref{Eq.MstarHalbach})) with respect to the ratio of the radii and equating it to zero. The resulting equation is, as can be seen below, a polynomial equation of order $p+1$, which does not in general have a closed form solution for $p>3$: 
\begin{align}\label{Eq_diff_eqs}
1-p\left(\frac{R_i}{R_o}\right)^{p-1}+(p-1)\left(\frac{R_i}{R_o}\right)^{p+1} = 0 &\qquad p > 1 \nonumber\\
-1+p-p\left(\frac{R_i}{R_o}\right)^2+\left(\frac{R_i}{R_o}\right)^{p+1} = 0 &\qquad p\leq -1
\end{align}
For the case of $p=1$ the solution is given by $\frac{R_i}{R_o} = e^{- W(-2 e^{-2})/2 - 1}$ where $W$ is the Lambert W function. The argument of the Lambert W function is greater than $-1/e$, which means that the function is single-valued. Evaluated numerically this corresponds to the well known ratio of $\frac{R_o}{R_i} = 2.2185$, although only the numerical solution has previously been given in literature \cite{Abele_1990,Coey_2003}. For other values of $p$ Eq.~(\ref{Eq_diff_eqs}) can be solved numerically. The solution, i.e.\ the highest efficiency and corresponding ratio of the radii, is shown in Figs.~\ref{Fig_Optimal_ratio_p} and \ref{Fig_Maximum_Mstar_p} as a function of $p$. In order to verify the analytical results for the field, the value of $M^{*}$ for the optimal ratio of the radii was also calculated numerically using Comsol Multiphysics finite element software (FEM) and was found to match the analytical value.

For the case of $p=-1$, the Halbach cylinder is simply a uniformly magnetized circle or infinite rod. For this system the efficiency becomes:
\begin{align}
M^{*}=\frac{1}{4}\left(1 - \left(\frac{R_i}{R_o}\right)^2\right).
\end{align}
This clearly has a maximum for $R_i = 0$, which means there is no internal bore. In this configuration the efficiency is 0.25, which as noted is the maximum efficiency possible.

\section{Demagnetization effects}
When a given Halbach design is implemented using permanent magnets with a finite coercivity, the demagnetization field internally in the magnet becomes an issue that need to be considered. Inside the magnet, the magnetic field $\mathbf{H}$ is not in the same direction as the magnetization $\mathbf{M}$, and if the component of the magnetic field in the direction opposite to the magnetization exceeds the coercivity, $H_c$, of the magnetic material, the magnet will be demagnetized \cite{Katter_2005}, with severe implications for the produced magnetic field and therefore also for the magnetic efficiency of the magnet design. Therefore it is important to consider demagnetization effects together with the magnetic efficiency.  For the $p=1$ Halbach it is known that the  reverse component of the magnetic field can exceed the intrinsic coercivity in regions around the inner equator \cite{Bjoerk_2008,Bloch_1998}, but for the general $p$ Halbach the demagnetization field is not known.

Using the expressions for the field inside the magnet from Ref. \cite{Bjoerk_2010a} it is straightforward to calculate the demagnetization of a Halbach cylinder. The condition for demagnetization to occur is
\begin{equation}\label{demag}
  \mu_0\frac{\mathbf{H}\cdot\mathbf{M}}{B_\n{rem}} \leq -H_c.
\end{equation}
We consider the cases $p=1$ and $p>1$ separately. The case of $p<1$ is not considered, but the calculations follow the case of $p>1$.

\subsection{Demagnetization in a $p=1$ Halbach}
The vector field inside the Halbach cylinder is given by \cite{Bjoerk_2010a}
\begin{align}\label{halbach1}
    A_z(r,\phi) & = B_\n{rem}(r \ln R_o-r\ln r)\sin \phi.
\end{align}
From this equation the magnetic flux density is readily derived in the same way as in Ref. \cite{Bjoerk_2010a}, and using elementary trigonometric relations we get
\begin{equation}\label{demagp1}
  \mu_0\frac{\mathbf{H}\cdot\mathbf{M}}{B_\n{rem}} = \mu_0^{-1}B_\n{rem}\left((\ln\frac{R_o}{R_i}-\frac{1}{2})\cos 2\phi -\frac{1}{2}\right),
\end{equation}

The condition for demagnetization becomes
\begin{equation}\label{demagcondp1}
  -\frac{\mu_0H_c}{B_\n{rem}}+\frac{1}{2} > (\ln\frac{R_o}{r}-\frac{1}{2})\cos 2\phi.
\end{equation}
Values of $(r,\phi)$ which fulfil the above inequality corresponds to the places in the Halbach cylinder where demagnetization will occur. As long as the prefactor on the cosine is positive, demagnetization will first arise for $\cos 2\phi=-1$, i.e.\ $\phi = \pm \pi/2$. This is on the `equatorial plane' of the Halbach cylinder. If the left hand side of Eq.~(\ref{demagcondp1}) is positive, i.e.\ if $\mu_0 H_c/B_\n{rem}<1/2$, then for any value of the inner radius $R_i$ there will be demagnetization.

Let us then consider $\mu_0 H_c/B_\n{rem}>1/2$. Demagnetization will always occur first at $r=R_i$. As $R_i$ is decreased there will be a critical value for which demagnetization first occurs on the inside of the Halbach cylinder. This critical value is given by
\begin{equation}\label{crit}
 \frac{\mu_0 H_c}{B_\n{rem}} = \ln\frac{R_o}{R_{m,\mathrm{crit}}},
\end{equation}
or
\begin{equation}\label{critr}
  R_{m,\mathrm{crit}} = \exp\left(-\frac{\mu_0H_c}{B_\n{rem}}\right)R_o.
\end{equation}
This is exactly equal to the value of the internal field produced by the Halbach cylinder. Demagnetization will thus first occur where the magnetization is pointing in the opposite direction of the field in the bore, and the value of the demagnetization field will be the Halbach field due to field continuity across the border between magnet and bore.

If the prefactor $\ln(R_o/r)-1/2$ is negative, demagnetization will first occur for $\cos 2\phi=1$, corresponding to $\phi = 0$ or $\pi$. In this case, demagnetization will occur from the critical radius:
\begin{equation}
  \frac{\mu_0 H_c}{B_\n{rem}} = 1-\ln\frac{R_o}{R_{\mathrm{crit}}}.
\end{equation}
or
\begin{equation}\label{critneg}
R_{\mathrm{crit}} = \frac{R_o}{\exp\left(1-\frac{\mu_0 H_c}{B_\n{rem}}\right)}.
\end{equation}
to $R_o$, with the largest value occurring at $r=R_o$, where the demagnetizing field will always have a value of $B_\n{rem}$. 

\begin{figure}[t]
  \centering
  \includegraphics[width=0.8\columnwidth]{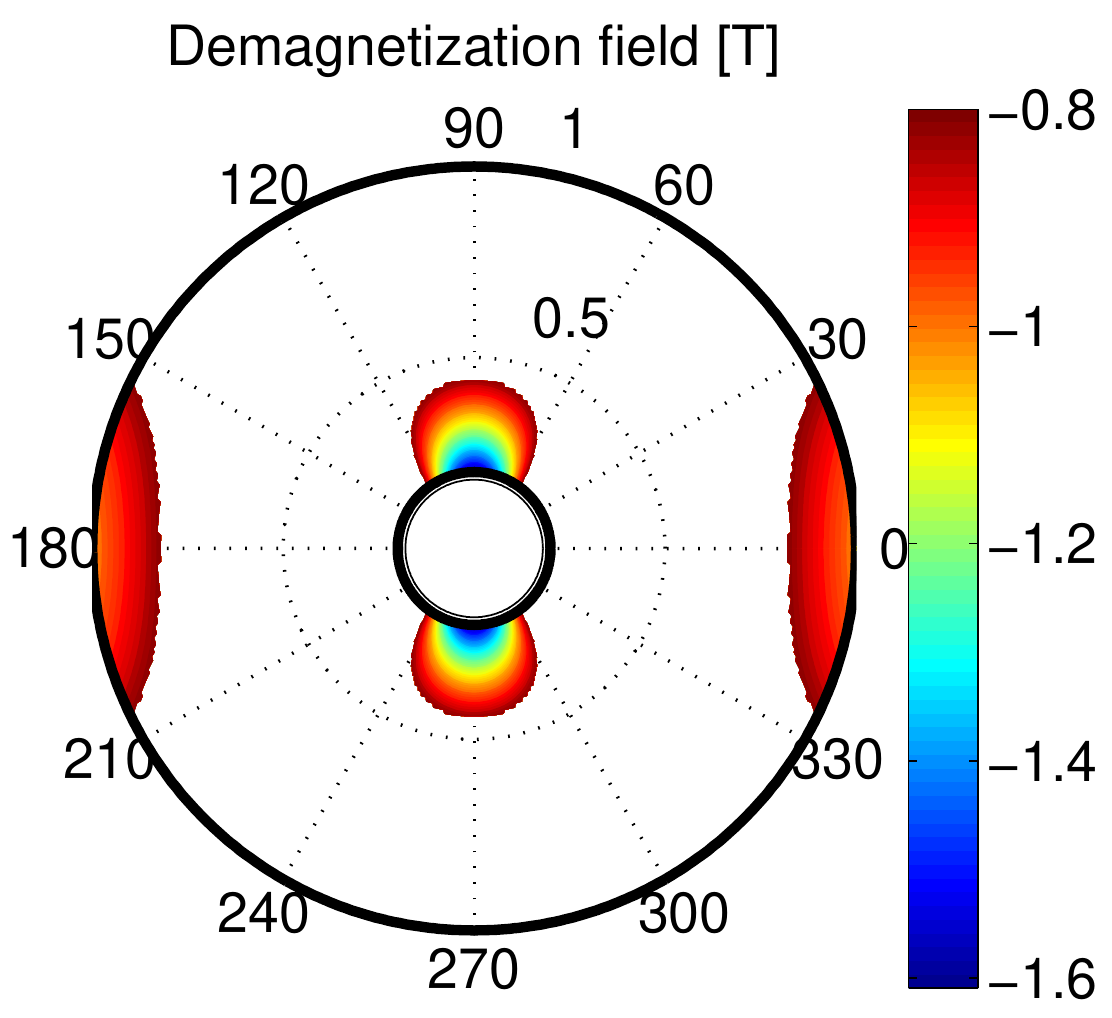}
  \caption{A polar plot of the demagnetization field for a $p=1$ Halbach cylinder with $R_o=1$, $R_i=0.2$ and $B_\n{rem}=1$ T for a coercivity of $\mu_{0}H_\n{c}=0.8$ T.}
  \label{Fig_Demag}
\end{figure}

Thus, for $1/2 < \mu_0 H_c/B_\n{rem}<1$ demagnetization can occur first both at $\phi = \pm \pi/2$ and at $\phi = 0,\pi$, while for $\mu_0 H_c/B_\n{rem}>1$ demagnetization always occur at $\phi = \pm \pi/2$ first. In other words, the demagnetizing field is always $B_\n{rem}$ at $\phi = 0,\pi$ and $r=R_o$ while it is $B_\n{rem}\ln\frac{R_o}{R_{m}}$ at $\phi = \pm \pi/2$ and $r=R_i$ and these are the maximum values of the demagnetizing field and thus to avoid demagnetization the coercivity must be larger than these values. An illustration of the demagnetization field for a $p=1$ Halbach cylinder with $R_o=1$, $R_i=0.2$ and $B_\n{rem}=1$ T for a coercivity of $\mu_{0}H_\n{c}=0.8$ T is shown in Fig. \ref{Fig_Demag}.

\subsection{Demagnetization in a $p>1$ Halbach cylinder}
The calculation of the demagnetization for $p>1$ proceeds very similarly to the case of $p=1$. Using the equations of the internal field in the magnet from Ref. \cite{Bjoerk_2010a} we get that the condition for demagnetization to occur becomes
\begin{equation}\label{demagpgt1}
  -\frac{\mu_0H_c}{B_\n{rem}}+\frac{1}{2} > \left(-\frac{p}{p-1}\left(\frac{r}{R_o}\right)^{p-1}+\frac{1}{2}\frac{p+1}{p-1}\right)\cos 2p\phi.
\end{equation}
When the prefactor of the cosine is positive (i.e.\ when $r<R_o(1/2+1/2p)^{1/(p-1)}$), demagnetization arises first for \\$\cos 2p\phi=-1$. Again, there will always be demagnetization for $\mu_0 H_c/B_\n{rem}<1/2$ no matter the value of $R_i$. For $\mu_0 H_c/B_\n{rem}>1/2$, the critical inner radius is given by
\begin{equation}\label{critrpgt1}
  \left(\frac{R_{m,\mathrm{crit}}}{R_o}\right)^{p-1} = 1 - \frac{p-1}{p}\frac{\mu_0 H_c}{B_\n{rem}}.
\end{equation}
Only for materials with an $H_c$ small enough that the right hand side of Eq.~(\ref{critrpgt1}) is positive, can demagnetization occur. It is evident that higher $p$ cylinders are less susceptible to demagnetization.

For $r>R_o(1/2+1/2p)^{1/(p-1)}$ demagnetization first arises for $\cos 2p\phi =1$, and in this case the demagnetization from $R_o$ to the critical radius is
\begin{equation}
  \left(\frac{R_{\mathrm{crit}}}{R_o}\right)^{p-1} = \frac{1}{p}+\frac{p-1}{p}\frac{\mu_0H_c}{B_\n{rem}}.
\end{equation}
This can only occur if $1/2 < \mu_0 H_c/B_\n{rem}<1$.

Similarly to the $p=1$ case, for $p>1$ there will always be critical demagnetization inside the Halbach cylinder for $\mu_0 H_c/B_\n{rem}<1/2$. For $1/2 < \mu_0 H_c/B_\n{rem}<1$ the first demagnetization can occur both for $\cos 2p\phi=1$ and $\cos 2p\phi=-1$, while for $\mu_0 H_c/B_\n{rem}>1$ demagnetization always occur first at $\cos 2p\phi=-1$. If $\mu_0 H_c/B_\n{rem}> p/(p-1)$ there will be no demagnetization at all in a $p$ Halbach cylinder.

The demagnetization field calculated above is for Halbach cylinders with a continuously rotating remanence. In most practical applications a Halbach cylinder is segmented, which can cause demagnetization to occur also at e.g. the interfaces between segments \cite{Bjoerk_2008}. However, the demagnetization field as calculated above will still generally be present and must be taken into account also for segmented designs. Several Halbach cylinder-like high field systems ($>3$T) have been constructed using permanent magnets \cite{Kumada_2001a,Kumada_2001b,Kumada_2003}. Here demagnetization is to some extend avoided by replacing magnets subject to a high demagnetization field with a high permeability soft magnetic material. However, the constructed magnets still produce a lower flux density than expected, which can be caused by the demagnetization issues discussed above.

\section{Discussion and conclusion}
Here we have calculated the magnetic efficiency of a general $p$-Halbach cylinder and showed that the efficiency decreases for increasing absolute value of $p$. We have also found the optimal ratio between the inner and outer radius, i.e. for the most efficiency design, as function of $p$. The optimal ratio cannot be expressed in closed analytical form, but the numerical solution is given. Finally, we have consider the demagnetization effect in a general $p$-Halbach cylinder, and showed that demagnetization is largest either at $\cos 2p\phi=1$ or $\cos 2p\phi=-1$. For a $p=1$ Halbach cylinder the maximum values of the demagnetizing field are at $\phi = 0,\pi$ and $r=R_o$ where the field is always $B_\n{rem}$, while it is the magnitude of the field in the bore at $\phi = \pm \pi/2$ and $r=R_i$. Thus to avoid demagnetization the coercivity must be larger than these values or the magnets at these locations must be replaced by a high permeability soft magnetic material.

\end{document}